\documentclass[aps,superscriptaddress,twocolumn,10pt]{revtex4-1}

\usepackage{amsmath,amsfonts,amssymb,graphicx,color,times,bbm,psfrag,dsfont,
slashed}
\usepackage[colorlinks=true,citecolor=blue]{hyperref}

\newcommand{\be}{\begin{equation}}
\newcommand{\ee}{\end{equation}}

\newcommand{\ha}{\frac{1}{2}}

\usepackage{bm}

\usepackage[normalem]{ulem}
\usepackage[T1]{fontenc}
\usepackage[latin9]{inputenc}

\newcommand{\beq}{\begin{equation}}
\newcommand{\eeq}{\end{equation}}
\newcommand{\bea}{\begin{eqnarray}}
\newcommand{\eea}{\end{eqnarray}}

\begin{document}

\title{Synthesis of Majorana mass terms in
low-energy quantum systems}

\author{L. Lepori}
\email[correspondence at: ]{llepori81@gmail.com}
\affiliation{Dipartimento di Scienze Fisiche e Chimiche, Universit\`a dell'Aquila, via Vetoio,
I-67010 Coppito-L'Aquila, Italy.}
\affiliation{INFN, Laboratori Nazionali del Gran Sasso, via G. Acitelli,
22, I-67100 Assergi (AQ), Italy.}

\author{A. Celi}
\affiliation{Institute for Quantum Optics and Quantum Information,
Austrian Academy of Sciences, Innsbruck, Austria.}
\affiliation{ICFO-Institut de Ciències Fotòniques, The Barcelona Institute of Science and Technology, 08860 Castelldefels (Barcelona), Spain.}

\author{A. Trombettoni}
\affiliation{CNR-IOM DEMOCRITOS Simulation Center, via Bonomea 265, I-34136 Trieste, Italy.}
\affiliation{SISSA and INFN, Sezione di Trieste,
via Bonomea 265, I-34136 Trieste, Italy.}

\author{M. Mannarelli}
\affiliation{INFN, Laboratori Nazionali del Gran Sasso, via G. Acitelli,
22, I-67100 Assergi (AQ), Italy.}

\begin{abstract}
We discuss the problem of how 
Majorana mass terms {\color{black} can be generated} in low-energy systems. 
We show that, while these terms imply the Majorana condition, the opposite is not always true when more 
than one flavour is involved. This is an important aspect for the  
low-energy realizations of the Majorana mass terms exploiting superfluid pairings, because in this case the Majorana condition is 
not implemented in the spinor space, but in an internal 
(flavour) space. Moreover, these mass terms 
generally involve opposite effective chiralities, 
similarly to a Dirac mass term. The net effect of these features
is that the Majorana condition does not imply a Majorana mass term. 
Accordingly
the obtained Majorana spinors, as well as
the resulting symmetry breaking pattern and low-energy spectrum,
are qualitatively different from the ones known in particle physics. 
This result has 
important phenomenological consequences, e.g. implies 
that these mass terms 
are unsuitable to induce an effective see-saw mechanism, 
proposed to give mass to neutrinos. 
Finally, we introduce and discuss schemes  
based on space-dependent pairings with nonzero total momentum 
to illustrate how genuine Majorana mass terms 
may emerge in low-energy quantum systems.
\end{abstract}

\maketitle


\section{Introduction} 
Due to their 
crucial role in physics beyond the Standard Model, a 
huge amount of research and interest is
devoted to the study and to the detection of Majorana fermions at 
CERN and in underground experiments. Majorana fermions were first introduced in 1937 by E. Majorana as real solutions of the Dirac equation~\cite{Majorana:1937}. The original motivation of Majorana was to prevent the existence of negative energy solutions. The resulting  fermionic particles coincide with 
their own antiparticles, then they are invariant under charge conjugation~\cite{Pal:2010ih,Aste:2008dc}
and neutral with respect to any additive charge~\cite{sakurai1967advanced}. The neutrality is encoded in the so-called Majorana condition, reading for a single-flavour relativistic fermion
\be
\psi  = C \, \psi^* 
\label{MNC}
\ee
(apart from a global phase), where $\psi$ is the real-space spinor and   the charge conjugation operator $C$ acts on the (suppressed) spinor indices.

 Closely related to the Majorana condition is the concept of Majorana mass.
If the $(3+1)$-dimensional Dirac equation has no mass term 
(Weyl equation), 
then the two (left and right) chiralities decouple. There are only two mass terms compatible with Lorentz invariance: 
the Dirac and the Majorana ones. Both terms couple spinors with opposite chiralities. However, the Dirac mass term
couples independent spinors, while the Majorana one couples chiralities related  by charge conjugation. 
A Majorana mass implies the fulfillment of~\eqref{MNC} 
\cite{Pal:2010ih,Aste:2008dc}.

Equation \eqref{MNC} with $C = {\bf 1}$ is also fulfilled by 
zero-energy excitations \cite{Wilczek:2009ul,Wilczek:2014uqa,Wilczek:2014lwa} 
(also dubbed Majorana modes) occurring at the edges of nontrivial topological 
insulators \cite{0034-4885-75-7-076501}. 
However, these excitations differ from  Majorana spinors because 
they lack of the internal spinor structure and 
do not obey fermionic statistics, but anyonic \cite{stern2008}. 
Majorana modes, as well as the simulation of the Majorana equation 
\cite{solano2011}, are not subject of investigation in this manuscript.  

Majorana masses provide a natural mechanism 
to give mass to the neutrino 
(required to explain oscillations \cite{petcov1987},
see \cite{Lan2011} for an ultracold atom simulation),  possibly
without introducing sterile right-handed chiralities for it
\cite{petcov1987,drewes}. 
Within the known Standard Model (SM) particles, neutrinos are the unique 
possible Majorana spinors;  remarkably only the Majorana mass  for the right-handed neutrinos is compatible with the symmetries of the SM~\cite{peskin1995introduction,Giunti:2007ry}. 
In the supersymmetric extensions of the SM 
\cite{doi:10.1142/9789812839657_0001}, 
a pletora of Majorana elementary particles is required, e.g.
as partners of  bosonic gauge fields. 
These Majorana particles are candidates to solve the long-standing problem of the 
dark matter component of the Universe~\cite{GellMann:1980vs, Yanagida:1979as, weinberg2000quantum}. 
In spite of these theoretical motivations, 
whether elementary Majorana particles exist in Nature is still an open question.
No evidence has been found
so far in running experiments, as in the neutrinoless double beta decay \cite{DellOro:2016tmg} and at LHC. The theoretical implications of the Majorana masses and 
spinors, as well as the perspective of observing elementary Majorana particles in extremely sensitive experiments, 
make desirable to obtain them in  low-energy quantum systems. 
For this purpose, it is crucial to identify  analogies and 
differences in these two frameworks. 

As we clearly show, there  are  important differences between Majorana fields
emerging in most of superfluid states
of metals and semimetals and the Majorana spinors 
defined in particle physics/{\color{black} high-energy systems}. 
For instance,  in the proposals considered in~\cite{Wilczek:2014lwa}, 
the fermionic pairing does not induce {\color{black} ``genuine'' -- in the sense 
of particle physics --} Majorana masses {\color{black} since the Majorana 
condition is not implemented in the spinor space, as in \eqref{MNC}, 
but in the flavour space.}

{\color{black} The main goal of the present paper is to 
identify and discuss the mechanisms for the emergence of genuine Majorana masses in low-energy models. Our key point originates from the observation 
that the Majorana condition does not imply the presence of mass terms 
for multi-flavour systems. To illustrate anyway 
the possibility of having the Majorana condition realized in the
{\em spinor} space we discuss schemes 
obtained exploiting unconventional superfluid pairings 
with nonzero total momentum.}
We finally present a Majorana mass inducing a Majorana condition 
where $C$ acts on both  spinor and flavour indices.


\section{General aspects of mass terms for spinors}
 We first consider the general structure of the mass terms for relativistic fermions (spinors), with particular emphasis on the associated symmetry breaking patterns. 
For simplicity, we neglect any interaction mediated by gauge bosons~\cite{Georgi:1985kw}.

For the sake of generality, we consider a fermionic system  of $N$  different flavors in $(3+1)$ dimensions, described by the Lagrangian 
$ {\cal L} = {\cal L}_K + {\cal L}_\text{mass}$, with ${\cal L}_K =i\, \sum_\alpha\bar\psi_\alpha\, {\gamma^\mu \partial_\mu} \,\psi_\alpha$, with $\alpha=1,\dots, N$  the  flavor index.
The Lagrangian ${\cal L}_K$   in the basis of Weyl (massless) spinors with definite chiralities $L , R$,  $\psi^t_\alpha=(\psi_{\alpha L}^t\,,\psi_{\alpha R}^t)$, \cite{Georgi:1985kw} is manifestly invariant under the product of unitary transformations{ $G = U(N)_{L} \times U(N)_{R}$
\cite{peskin1995introduction,Georgi:1985kw}.} 

Any mass term cannot entirely preserve $G$; 
the most general Lorentz invariant 
one~\cite{Pal:2010ih,Aste:2008dc,petcov1987} can be written as
$
 {\cal L}_\text{mass} =  \sum_\alpha \left( 
{\cal L}_{\text{Lm},\alpha} + {\cal L}_{\text{Rm},\alpha} + 
{\cal L}_{\text{Dm},\alpha} \right)$, 
with 
$ {\cal L}_{\text{Lm},\alpha} =  \frac{m_L}2 {\psi}^t_\alpha \gamma_0 \, C^\dagger P_L{\psi}_\alpha + \text{H.c} $, 
 ${\cal L}_{\text{Rm},\alpha}  = \frac{m_R}2{\psi}^t_\alpha \gamma_0 \, C^\dagger P_R {\psi}_\alpha + \text{H.c}$,
 $  {\cal L}_{\text{Dm},\alpha}  = - m_D \bar {\psi}_\alpha  {\psi}_\alpha +  \text{H.c} $,
where $P_{L/R}=(1\mp \gamma_5)/2$. For simplicity we have assumed that all the  flavors have equal masses. 
The matrix $C= i \gamma^2$, in the Weyl basis reduces to  $C = i \sigma_2 \otimes  i \sigma_2$, {\color{black} thus, in} terms of Weyl spinors, we obtain
\begin{align}
 {\cal L}_{\text{Lm},\alpha}  &= - \frac{m_L}2  \psi_{\alpha L}^t i\sigma_2 \psi_{\alpha L} + \frac{m_L}2 \psi_{\alpha L}^\dagger i\sigma_2 \psi_{\alpha L}^*,  \label{eq:mL2} \\
  {\cal L}_{\text{Rm},\alpha}  &=  \frac{m_R}2 \psi_{\alpha R}^t i\sigma_2 \psi_{\alpha R} -\frac{m_R}2 \psi_{\alpha R}^\dagger i\sigma_2 \psi_{\alpha R}^*  \label{eq:mR2}\,, \\
   {\cal L}_{\text{Dm},\alpha}  &= - m_D \, \psi_{\alpha L}^\dagger  \psi_{\alpha R} -m_D \, \psi_{\alpha R}^\dagger  \psi_{\alpha L} \, .  \label{eq:mD2}
\end{align}

The Dirac mass  in
\eqref{eq:mD2} mixes the chiralities, locking left- and right-handed 
chiral rotations. 
The resulting  breaking pattern is
$ G \to SU(N)_{V} \times U(1)$,
where  $SU(N)_{V}$ and $U (1)$ involve the same (simultaneous) transformations on the $L$ and $R$ spinors (see for example~\cite{Georgi:1985kw,lepori2015}).

The terms in \eqref{eq:mL2}-\eqref{eq:mR2} break the 
number symmetry $U(1)$ but {\it do not mix} the $L$ and $R$ chiralities 
(indeed, starting from a chirality, the opposite one is obtained by charge 
conjugation), leading to 
$ G \to O(N)_{L} \times O(N)_{R}$, where $O(N)_{L/R} \subset U(N)_{L/R}$ are orthogonal groups. 

The dispersion laws corresponding to \eqref{eq:mL2}-\eqref{eq:mD2} are
$E_{\pm} = \sqrt{|\bm{p}|^2 +m_{\pm}^2}$,
with
\be
\label{eq:eigenmasses}
m_\pm = \ha  \left\vert m_L+m_R \pm \sqrt{(m_L-m_R)^2+4 m_D^2}\right\vert \,.
\ee
This mass splitting is of the utmost phenomenological importance, 
because it allows for the generation of a massive left neutrino through the see-saw mechanism, see for example~\cite{petcov1987,Giunti:2007ry}. 
As an aside, we note that  \eqref{eq:mL2}-\eqref{eq:mD2} do not allow any phase redefinition of $m_R$, $m_L$, and $m_D$. 
Therefore, if one of these masses acquires a complex phase, the product CP of charge and parity conjugation symmetries is broken, see for example \cite{weinberg2000quantum, petcov1987}.  Instead, if $m_D = 0$, the relative sign between $m_L$ and $m_R$, that can be difficult to set in low-energy simulations, is re-absorbable, then unphysical.

\subsection*{Majorana condition versus Majorana mass terms}
 Importantly, a Weyl  spinor (say with  chirality $L$ and flavour $\alpha$),  acquiring a Majorana mass, gives rise  to the Majorana spinor, $\psi_{\alpha M}^t = (\psi_{\alpha L}, -i \, \sigma_2 \psi_{\alpha L}^*)$, fulfilling~\eqref{MNC}.
{\color{black} However}, the fulfillment of a Majorana condition does not necessarily imply the presence 
of a Majorana mass  if $N \neq 1$.  Indeed, in this case the same condition can be realized on the flavour indices: $\tilde{\psi}_{\alpha}  = \tilde{C}_{\alpha \alpha^{\prime}} \, \tilde{\psi}_{\alpha^{\prime}}^*$. The symbols $\tilde{\psi}_{\alpha}$ denote fermionic fields, even not relativistic, where chiralities can be unspecified (or even not defined) in general; moreover $\tilde{C} \neq C $ typically. This Majorana condition can be related with mass terms reading as 
\beq
\tilde{\psi}^{\dagger}_\alpha  \, \tilde{C}_{\alpha \alpha^{\prime}} \tilde{\psi}^*_{\alpha^{\prime}}  + \mathrm{H. \, c.} \, ,
\label{MC2}
\eeq
 explicitly breaking the Lorentz invariance and mixing in general all the chiralities (if defined). This situation is largely encountered in low-energy physics and represents an important obstruction against the realization of genuine Majorana masses; explicit examples will be given in the following. {\color{black} Finally, situations where the Lorentz invariance is broken by definite flavour structures exist also in the context of neutrino physics \cite{blasone1995,blasone2005}.}


\section{Weyl spinors on lattice systems}   
Weyl spinors, the starting building blocks for the mass terms, can emerge as low-energy excitations in condensed matter three-dimensional (3D) systems \cite{savrasov2011,Lv:2015yu,Lv:2015kx,Xu613,Xu294,Xu:2015nr,Lu622},  called Weyl semimetals. 
 Notably, they host two inequivalent {\color{black} and isolated points (Weyl nodes) in the Brillouin zone where two bands touch each others. These points} are separated in momentum space,  breaking the  spatial inversion or time reversal canonical symmetries \cite{ludwig2009,lepori2015w}.
Close to the Weyl nodes, the fermionic quasiparticles have  a linear dispersion law and their dynamics can be effectively described by two Weyl Hamiltonians with definite chiralities. What differentiates the various models is the shape of the Brillouin zone and the momentum separation between the Weyl nodes. 
Instead, their appearance  in pairs has a topological origin \cite{Nielsen:1981hk,Rothe:1992nt}. 

Some 2D bipartite lattice models (dubbed naive Dirac semimetals), not breaking chiral symmetry \cite{Nielsen:1981hk,Rothe:1992nt} and still hosting isolated band-touching points, can be also thought as 3D Weyl semimetals. In this set are
the honeycomb lattice \cite{Wallace:1947,semenoff1984} (characterizing graphene \cite{CastroNeto:2009zz}), the brick-wall lattice (recently realized experimentally \cite{Tarruell:2012zz}), and the square lattice pierced by a magnetic $\pi$-flux per plaquette \cite{Affleck:1987zz}. Indeed, these 2D models (connected by an interpolating pattern~\cite{Mazzucchi:2013}) are also related with genuine Weyl semimetals by a projection along one axis.   {\color{black} Reversely, by stacking the former models and adding  suitable tunnelings along the stacking direction, one can obtain the latter ones} ~\cite{Laughlin:1990,Lepori:2010rq,burkov2011,burrello2017} (in this way, 
 anisotropic and non-linear dispersions can be also obtained \cite{bernevig2012,huang2016,lepori2016}).

Also motivated by the {\color{black} previous} discussion, for our purposes we focus primarily on the honeycomb lattice described by a tight-binding Hamiltonian ${\cal H}_{\mathrm{hon}}$ with spectrum $\epsilon(\bm{k})$ \cite{Wallace:1947,semenoff1984,CastroNeto:2009zz}. Expanding ${\cal H}_{\mathrm{hon}}$ around the Weyl nodes at $\bm{k}_R$ and  $\bm{k}_L$, up to a unitary transformation, we  obtain the Weyl Hamiltonian \cite{Wallace:1947,semenoff1984,Affleck:1987zz} 
 \be
{\cal H}_{\mathrm{LE}} (\bm{p})= 2 t \, \sum_{\alpha} \, \int d\bm{p} \, \Big( \psi^{\dagger}_{\alpha R}(\bm{p}) \, \bm{\sigma} \cdot \bm{p} \; \psi_{\alpha R} (\bm{p}) \, -  \, \big(L\big) \Big) \, , 
\label{ham3D}
\ee
where $\bm{p} = \bm{k}- \bm{k}_{R,L}$, with  $|\bm{p}| \ll |\bm{k}_{R/L}|$ the residual momentum,  $t \equiv 1/2$ is the tunneling amplitude,
$\epsilon (\bm{k}_{R,L} + \bm{p}) \approx  |\bm{p}|$,
and $\psi_{\alpha R/L}(\bm{p}) = \big(c_{A, \alpha} \big(\bm{k}_{R/L} + \bm{p}) , c_{B, \alpha} (\bm{k}_{R/L} + \bm{p}\big)\big) $, with $c_{A, \alpha}$ ($c_{B, \alpha}$)  annihilation operators acting on the $A$ ($B$) sublattice. 
 
The Hamiltonian \eqref{ham3D} describes the low-energy physics also  for all 
the other semimetals mentioned above. In the following, it will be chosen  as the starting point for the implementation of the different mass terms in \eqref{eq:mL2}-\eqref{eq:mD2}.


\section{Majorana-like masses}  Mass terms as in \eqref{MC2} are obtained by appropriate attractive interactions between fermions in a metal or a (Weyl) semimetal, turning it into a superfluid, see for example~\cite{Wilczek:2009ul,Wilczek:2014uqa,Wilczek:2014lwa}. This general fact can be understood 
considering, as a leading example, an onsite interaction 
$- \, U \, \sum_{i} \, c_{i, \uparrow}^{\dagger} c_{i, \uparrow} \,  c_{i,\downarrow}^{\dagger} c_{i,\downarrow}$,   $U >0 $, between two flavours $\{\uparrow , \downarrow\}$,
and  defining  the  two-spinor
$\Phi (\bm{k}) = \big(c_{\uparrow} (\bm{k}), c_{\downarrow} (\bm{k}) \big)$. 
The resulting mean field BCS term is
\be
{\cal H}_\text{BCS} (\bm{k}) = \Delta \, 
\Big(\Phi^{\dagger} (\bm{k}) \,  i \sigma_2 \, \Phi^* (-\bm{k}) \, + (\bm k \to - \bm k) \Big) + \text{H.c.} \,,
\label{majomass0}
\eeq
formally similar to \eqref{eq:mL2}-\eqref{eq:mR2}. 
The emergence of a field fulfilling a Majorana condition
can be made explicit defining the {\color{black} field
 $\Psi (\bm{k}) = \Big(\Phi (\bm{k}) , 
- i \sigma_2 \, \Phi^* (-\bm{k}) \Big)^{T}$}
and expressing ${\cal H}_\text{BCS} (\bm{k})$ in terms of it.  
The appearance of $\Psi (\bm{k})$ is deeply related with the presence of both positive- and negative-energy solutions of the Bogoliubov-de Gennes equations~\cite{annett2004} connected  by $C = \sigma_2 \otimes \sigma_2 $, since for the total Hamiltonian ${\cal H}(\bm{k})$ it holds ${\cal H}(\bm{k}) = - \, C^{-1} \, {\cal H}^*(- \bm{k}) \, C$, see for example~\cite{ludwig2009} and references therein. This is a general feature of  superconducting systems, even if the form of $\Psi (\bm{k})$ can vary, depending for instance on the number of flavours or lattice indices.

Assuming now to work on a Weyl semimetal (the discussion above still {\color{black} applies}, since we can neglect the sublattices indices), we examine in more detail the chiral structure of the superfluid term
\eqref{majomass0}. To this end, we expand it close to the Weyl points, {\color{black} obtaining a pairing Hamiltonian}
\beq
{\cal H}_\Delta = - {\color{black} \Delta}   \int\! d \bm{p}  \Big(  \Phi_{R }^{\dagger} (\bm{p}) \,  i \sigma_2 \, \Phi_{L}^* (-\bm{p}) +  \, R  \rightleftarrows L \, +  \text{H.c.} \Big) \,,
\label{majomass}
\eeq
clearly showing that this mass term does not induce the breaking pattern of a Majorana mass,
because it
couples quasiparticles with opposite 
chiralities (momenta), as a Dirac mass. Therefore, {\color{black} despite being 
a {\em Majorana-like} mass}, this is not
a {\em genuine} Majorana mass. 
For the corresponding low-energy spectrum, in the simultaneous presence of a Dirac mass, we obtain (at vanishing chemical potential)
\beq
\lambda_{\mathrm{MD}} (\bm{p}) =  \sqrt{|\bm p|^2+m_D^2 +  \Delta^2 } \,,
\eeq
which does not coincide  with the one in \eqref{eq:eigenmasses}.

Another central difference  is that the matrix $i \sigma_2$  in  \eqref{majomass0}-\eqref{majomass} acts on the flavour space, as in \eqref{MC2}, and not on the spinor (sublattice) indices  as in  \eqref{eq:mL2}-\eqref{eq:mR2}. Notice that the same crucial difference allows to define a Majorana field  also in superfluid phases of ordinary metals, where the Fermi surface is extended and no effective chiral spinors occurs.


\section{Genuine Majorana masses} From the previous discussion, {\color{black} 
it emerges that engineering a genuine Majorana mass (and the corresponding symmetry breaking pattern) by suitably coupling the nodal points of a Weyl semimetal,  
necessarily requires the implementation of the charge conjugation operation, as in \eqref{MNC}, in the spinor (sublattice) indices.
Moreover, it requires a superfluid pairing in single chiral valleys, $\bm{k}_L$ or $\bm{k}_R$, then with nonzero total momentum. 
 
We conclude that the request to implement 
genuine Majorana mass is to have  intra-valley couplings, still enforcing 
the Majorana condition on the sublattice indices. Candidates to 
realize such pairings are naturally Weyl semimetals, 
as the ones obtained from both spinless or spinful non-relativistic fermions in honeycomb lattices, loaded up to near half filling, with suitably engineered two-body interactions. 
Indeed, in Weyl semimetals, specific interactions can induce 
spatially dependent pairings with nonzero total quasimomentum, 
that are analogous to the FFLO pairing in the continuous space~\cite{FF,LO,Anglani:2013gfu}, but are expected to be more robust against disorder than standard FFLO  (see for example~\cite{shi2012,moore2012}).
  
Let us start 
from the spinless case, where one has only one species of non-relativistic 
fermions on the lattice, giving rise to a single pair of Weyl spinors ($N=1$). 
In this case, the desired intra-valley superfluid pairing could be  
energetically favored by large 
nearest-neighbor (inter-sublattices) 
attractions, and possibly stabilized 
by a further (subleading) next-nearest-neighbor attraction \cite{Yao2015} 
(otherwise phase separation may prevent superfluidity~\cite{Orus:2012,Capponi:2017}). In ultracold atom experiments, 
the required nearest-neighbor interaction between fermions can be synthesized  
for instance as an effective interaction mediated by ($s$-wave) collisions with bosons (see for example~\cite{lim2009,massignan2009,Cirac2010}).
Another possibility would be to exploit dipolar interactions in 
fermionic magnetic atoms like Erbium, 
where stable dipolar Feshbach resonances between different spin states have been experimentally demonstrated very recently in \cite{baier2018realization}.} 

Assuming $i \in A$ and $j \in B$ nearest-neighbour, 
{\color{black} similarly as in \cite{sigrist2016},}
the direction-dependent spin-triplet superfluid term  
can be written as 
{\color{black}
\begin{align}
\langle c_{i  A} c_{j B}  \rangle =  \langle c_{i  A} c_{j  B} -  c_{j  B}  c_{i A}  \rangle \equiv \Delta_{i,j}  = \nonumber \\
{} \nonumber \\
= \Delta(\bm{k}_R) \,   e^{i \bm{k}_R \cdot (\bm{i} + \bm{j})} \pm \Delta(\bm{k}_L) \, e^{i  \bm{k}_L \cdot (\bm{i} + \bm{j})} \, ,
\label{pairspinless}
 \end{align}
 where we neglected a $\bm{p}$ dependence of the pairings, being $\pm 2 \, \bm{p}$ the relative momentum between the fermions in the pair.  Due to the Fermi statistics, which 
 constrains $T  +  J + L$ to be odd~\cite{uchoa2007} ($T, J$, and $L$ being the lattice, flavour, and angular quantum numbers, respectively), one finds that pairing functions in the two valleys are even in $\bm p$: 
 $\Delta(\bm{k}_{R/L} \bm{p}) = \Delta(\bm{k}_{R/L}, - \bm{p})$.}
 Notably, for each chiral valley,
 only one plane wave appears: this is a necessary condition  to have a spatially inhomogeneous pairing with a finite gap, see  \cite{Anglani:2013gfu} for an extended discussion.

If the condition \eqref{pairspinless} is satisfied, then close to $\bm{k}_L$ or  $\bm{k}_R$ the pairing  Hamiltonian reads
 \be
{\cal H}_{M} = \int \mathrm{d} \bm{p}  \, \Big(\Delta(\bm{k}_R) \, \psi_{R }^{\dagger} (\bm{p}) \,  i \sigma_2 \, \psi_{R}^* (-\bm{p}) - (R\to L)  \big)+  \text{H.c.} \Big) \, ,
\label{majomass2}
\ee
{\color{black} after a phase redefinition of the  $\psi_{L} (\bm{p})$ fields, required if the minus sign holds in \eqref{pairspinless} \cite{herbut2010,sigrist2016} (see also in the following); indeed this sign is
 unphysical if Dirac mass terms are not present at the same time, so that it can be reabsorbed.
 
The expression \eqref{majomass2} coincides with  \eqref{eq:mL2}-\eqref{eq:mR2}: now the matrix $\sigma_2$ acts}
on the spinor (sublattice) indices as desired. Therefore, two genuine Majorana masses are generated, involving the two chiralities separately, 
and realizing the corresponding breaking pattern.  At variance, the spinless $p$-wave pairing in \cite{poletti2011} also induces \eqref{MNC}, however opposite chiralities are paired, due to the zero momentum of the Cooper pairs, meaning that, {\color{black} in this case,} genuine Majorana masses are not generated. 

A similar {\color{black} mechanism works also for schemes based on
two-component non-relativistic fermions, 
leading to $N = 2$.} In this case, the required intra-valley pairing has been found  favored by various authors close to half filling (in the presence of a nearest-neighbours attraction and possibly of a subdominant onsite repulsion or attraction \cite{herbut2010,shi2012,moore2012,senechal2016}).
The relevant  (singlet or triplet) pairings are $\Delta_{i,j} \sim \langle c_{i A \uparrow} c_{j  B \downarrow}  \pm  c_{i A  \downarrow} c_{j B  \uparrow} \rangle $. 

For the triplet pairing in honeycomb lattices, {\color{black} equations \eqref{pairspinless}
and \eqref{majomass2} still hold, 
 with the replacement 
$ \psi_{R} (\bm{p}) \to \psi_{R, \alpha} (\bm{p})\equiv \big(c_{A, \alpha}(\bm{p}) , c_{B, -\alpha}(\bm{p}) \big)$, and the trace over flavour index $\alpha=\pm 1$ is taken.
Again, the matrix  $\sigma_2$ acts on the sublattice indices, instead the identity on the flavour space is understood. The triplet pairing is also called Kekule ansatz \cite{herbut2010,sigrist2016}; two configurations, $s$ and $p$, are possible for it, connected with the sign in \eqref{majomass2}.}

{\color{black} Let us briefly discuss about possible experimental 
setups in which the Majorana mass term can be synthesized in the $N=2$ case. 
Remarkably, a spin-triplet intra-valley pairing, enforcing 
the Majorana condition on the sublattice indices, 
has been experimentally found in $\mathrm{Cd}_3  \mathrm{As}_2$ crystals 
\cite{wang2016}, which display a semimetal behavior. A similar pairing can 
be also induced in ultracold atoms realizations of the Kane-Mele 
model \cite{KM}, a two-species variant of the 
Haldane model, the latter being experimentally realized in \cite{Jotzu:2014ek}:
one needs to add  a 
nearest-neighbor attractive interaction. Let us call $V$ the magnitude of such interaction. 
In \cite{sigrist2016}, for zero or negligible on-site interactions, the spin-triplet 
paired superfluid phase arises for $V$ larger than a critical value $V_c$, and it apparently 
persists also in the limit of vanishing spin-orbit coupling. For $V\ge V_c$, the spin-triplet 
order parameter $\Delta_t$ is an increasing function of $V$. 
To observe such superfluid phase, 
it is reasonable to expect that one has to 
achieve $\Delta_t$ larger than thermal excitations, thus a sufficiently large $V$ such that $\Delta_t\gtrsim k_B T$, where $k_B$ is the Boltzmann constant and $T$ the temperature of the sample. 
In ultracold atoms experiments, the two energy scales have not absolute meaning, but both depend on the band width (which also determines the Fermi energy), 
and can be expressed in units of the tunneling $t$. Indeed, on one hand, $t$ is an obvious energy scale for the lattice Hamiltonian and its interactions.
On the other hand, in ultracold atom experiments the key parameter is the achievable entropy per atom, which fixes
the value of $T$ to be some fraction of $t$, say 
$k_B T \equiv \nu \, t$. In state-of-the-art experiments with fermionic atoms, values 
$\nu \sim 0.25$  
are currently achievable \cite{mazurenko2017cold}. Assuming such values, from \cite{sigrist2016}(Fig. 9)  we see that $\Delta_t\gtrsim 0.25 t$ 
requires $V\gtrsim 3 t$. Such magnitudes for the nearest-neighbor interactions are also within the experimental reach, for instance through magnetic dipolar couplings. 
Indeed, similar magnitudes have been already  demonstrated experimentally, e.g. in bosonic erbium \cite{Baier2016}. 
The main challenge in the described scheme appears therefore to combine all the required ingredients 
in the same experiment.
}

For the singlet case \cite{moore2012}, Majorana masses can also be synthesized.
The intra-valley pairing,  for which $\Delta_s(\bm{k}_R , \bm{p}) =  \pm \Delta_s(\bm{k}_L , \bm{p})$ holds, induces a (modified) Majorana condition involving,
in the basis $\psi_{R/L,n,\alpha} = c_{n,\alpha} (\bm{k}_{R/L})$ ($n=1,2$ labeling the  sublattice $A$ and $B$), both the chiral and the flavour indices (a  situation also considered in  particle physics~\cite{petcov1987}), symmetrically. Indeed,
using the known relation $\epsilon_{n , m} \, \epsilon_{\alpha , \beta} = \delta_{m , \alpha} \delta_{n , \beta} - \delta_{m , \beta} \delta_{n , \alpha}$ , we obtain that
the pairing in real space $\Delta_{i,j} \equiv \Delta_s$ (independent of $n, m$)  can be written as
\beq
\Delta_s  = \epsilon_{m , n} \, \epsilon_{\alpha , \beta} \langle c_{m, \alpha} c_{n, \beta}\rangle = (i \sigma_2)_{m , n} \, (i \sigma_2)_{\alpha , \beta} \langle c_{m, \alpha} c_{n, \beta}\rangle \, ,
\eeq
and we obtain the low-energy Hamiltonian  (say close to $\bm{k}_R$) 
\beq
{\cal H}_{M,R} = \int \mathrm{d} \bm{p}  \, \Big(\Delta_s(\bm{k}_{R}) \, \psi_{R,m,\alpha}^{\dagger} (\bm{p}) \, M_{m,n, \alpha, \beta} \, \psi_{R, n, \beta}^* (-\bm{p})  \Big)  \,,
\label{singlet}
\eeq
with $M_{m,n, \alpha, \beta} = \big( (i \sigma_2) \otimes (i \sigma_2) \big)_{m, n, \alpha , \beta}$.
The Fermi statistics implies
$\Delta_s(\bm{k}_{R/L}, \bm{p}) = - \Delta_s(\bm{k}_{R/L}, - \bm{p})$, then $\Delta_s(\bm{k}_{R/L}, 0) = 0$:
 a vanishing pairing 
 occurs at the Weyl momenta (hidden order \cite{uchoa2007,herbut2010}), therefore, to obtain a stable pairing, the  atomic filling of the lattice assumes a more relevant role than in the triplet case.

In \eqref{pairspinless}-\eqref{singlet} we always set $|\Delta_{(s/t)}(\bm{k}_R , \bm{p})| = |\Delta_{(s/t)}(\bm{k}_L , \bm{p})|$, since in most of realistic systems the fermionic attractions are independent on the total momentum $\bm{K}$ of the interacting pair. However, an unbalance between the  pairings can be induced in  ultracold atomic mixtures  \cite{lim2009,massignan2009}, forcing the Bose-Bose  or the Fermi-Bose interaction to depend also on $\bm{K}$.  A recent proposal to achieve this dependence exploits a magnetic Feshbach resonance modulated by two Raman laser beams propagating along different directions, then exploiting the Doppler effect \cite{jie2017}.  This technique could also yield an additional  {\color{black} controllable parameter, beyond the filling and the interaction strengths,}   to favor the Majorana masses  \cite{he2017}. 


\subsection*{Simultaneous effect of Dirac(-like) masses} 
 On the honeycomb lattice, a further mass term
can be  synthesized  by an energy offset between the sublattices $A$ and $B$ \cite{Kogut1975,semenoff1984}
${\cal H}_\text{off} = M_\text{off} \, \Big( \sum_{\alpha , i \in A} \, c_{i, \alpha}^{\dagger} c_{i, \alpha} -  \sum_{\alpha , j \in B } \, c_{j, \alpha}^{\dagger} c_{j, \alpha} \Big),$
leading to 
${\cal H}_\text{off}= M_\text{off} \, \sum_{\alpha} \,  (\psi_{\alpha L}^{\dag} \,  \sigma_3 \, \psi_{\alpha L} + \, \psi_{\alpha R}^{\dag} \, \sigma_3 \, \psi_{\alpha R} )$, that is not a genuine Dirac mass. However,
the low-energy spectrum of  ${\cal H}_{\mathrm{hon}} + {\cal H}_{\mathrm{off}}$   reads
$\lambda_{\mathrm{D}} (\bm{p}) =  \sqrt{|\bm{p}|^2 +  M_{\mathrm{off}}^2 }$,
as for standard BCS superfluids \cite{annett2004}. 
But, when genuine Majorana masses $\propto m_{L/R} $ are also included,  the total spectrum reads
\beq
E_{\mathrm{hon} , \pm} = \sqrt{\big(|\bm{p}| \pm m_{L/R} \big)^2 + M_\text{off}^2} \, ,
\eeq
differing from \eqref{eq:eigenmasses}.
The reason of this mismatch is that ${\cal H}_\text{off}$ does not have the correct chiral structure. 

Let us now consider a different set-up, that is  the $\pi$-flux lattice, with free Hamiltonian ${\cal H}_K$\cite{Affleck:1987zz, Lepori:2010rq}. There, exploiting  a peculiar  periodicity of the magnetic Brillouin zone, a Dirac mass  can be achieved 
by a Bragg pulse scheme \cite{Lepori:2010rq}. 
This procedure, based on the continuous transfer between the Weyl points, effectively synthesizes the term ${\cal H}_\text{Bragg} 
= M_D \, \sum_{\alpha} \, 
 (\psi_{\alpha L}^{\dag}  \psi_{\alpha R} + \, \psi_{\alpha R}^{\dag} \psi_{L\alpha} )$ 
 in \eqref{eq:mD2} and still leads to $\lambda_{\mathrm{D}} (\bm{p})$, with $M_\text{off} \to M_D$. 
Now, if one includes the  Majorana masses, the resulting total spectrum coincides now with  ~\eqref{eq:eigenmasses}, {\color{black} provided that the minus sign holds in front of the left pairing in \eqref{pairspinless} and \eqref{majomass2}.}
Technically, the difference between the total spectra for the two lattices is due to the fact that  ${\cal H}_{\mathrm{hon}}  + {\cal H}_\text{off}$ and ${\cal H}_K + {\cal H}_{\mathrm{Bragg}}$, expanded close to the Weyl nodes, are not equal but only unitary equivalent~\cite{semenoff1984,Affleck:1987zz,CastroNeto:2009zz} (due to the different Pauli matrices appearing in \eqref{ham3D} in the two cases \cite{semenoff1984,Affleck:1987zz}). Indeed  ${\cal H}_\text{off}$ is not a genuine Dirac mass  as \eqref{eq:mD2}, since it does not mix the opposite chiralities. Therefore, although the two lattices share the same spectrum in the absence of  Majorana masses, they behave differently if the latter terms
are also considered.


\section{Outlook} Various extensions of the present work are in order, 
including  \emph{i)} 
the synthesis of Majorana spinors from a superfluid phase on the $\pi$-flux square (cubic) lattice, possibly via the same schemes working for the honeycomb;   \emph{ii)} the  detailed investigation of the simultaneous coexistence of Majorana and Dirac masses on the described Weyl lattices, also including fluctuations;  \emph{iii)} the realization of a Majorana mass in the topological Haldane model \cite{Haldane:1988zza}, recently  experimentally achieved \cite{Jotzu:2014ek}, and hosting at criticality a unique chiral node. { Finally, it would be interesting to study the  
Zitterbewegung \cite{sakurai1967advanced,qu2013,leblanc2013}, as a tool to discriminate Majorana and Dirac masses  $m_{D/M}$. Indeed the oscillation amplitudes 
are expected to differ in the two cases, due to the different spinor structures \cite{sakurai1967advanced}. In the described lattice set-ups, a first estimate of the amplitudes  is $\sim \frac{t}{m_{D/M}} \, a$, with oscillations of order of few lattice sizes.\\

\section*{Acknowledgements}
The authors are pleased to thank M. Burrello, D. Giuliano, E. Molinaro, and S. Paganelli for many useful discussions.
A. C. acknowledges the support of Spanish MINECO (SEVERO OCHOA Grant
SEV-2015-0522, FISICATEAMO FIS2016-79508-P)
the Generalitat de Catalunya (SGR 874 and CERCA program),
Fundaci\'o Privada Cellex, and EU grants EQuaM (FP7/2007-2013 Grant
No. 323714), OSYRIS (ERC-2013-AdG Grant No. 339106), QUIC (H2020-FETPROACT-2014 No. 641122), SIQS (FP7-ICT-2011-9 Grant No. 600645).
The authors also acknowledge a fruitful participation 
to the workshop "From Static to Dynamical Gauge Fields with Ultracold Atoms",
in the Galileo Galilei Institute for Theoretical Physics, Firenze, 22th May - 23th June 2017, where part of this work has been
performed.

\bibliography{BIB}
\end{document}